\documentstyle[11pt]{article}
\newcommand{\Si}{\Sigma}
\newcommand{\si}{\sigma}
\newcommand{\ove}{\overline}
\newcommand{\Sir}{\ove{\Sigma}}
\newcommand{\sir}{\ove{\sigma}}
\newcommand{\sirce}{\ove{{\sigma}}_{0}}

\newcommand{\sig}{<\!\sigma,\ove{\sigma}\!>}
\newcommand{\Sig}{\Sigma\times\ove{\Sigma}}
\newcommand{\cero}{<\!\sigma,\ove{\sigma}_{0}\!>}
\newcommand{\conu}[6]{[\hat{#1}_{#2},\hat{#3}_{#4}]=i\hbar\epsilon_{#2#4#5}
\hat{#6}_{#5}}
\newcommand{\cond}[4]{[\hat{#1}_{#2},\hat{#3}_{#4}]=0}
\newcommand{\cont}[3]{[\hat{#1}_{#2},\hat{#3}]=0}
\newcommand{\hateq}{\hat{=}}
\newcommand{\op}[2]{\hat{#1}_{#2}}
\newcommand{\lista}[2]{\newcounter{#1}\begin{list}
{$\bf #2_{\arabic{#1}}$}{\usecounter{#1}}}
\newcommand{\ee}{\]}
\newcommand{\be}{\[}
\newcommand{\hs}{\hspace*{1.6em}}
\newcommand{\teo}[2]{\hat{U}(#1,#2)}
\newcommand{\fa}{\forall}
\newcommand{\sa}{\hfill (SA)}
\newcommand{\bc}{\begin{center}}
\newcommand{\ec}{\end{center}}
\newcommand{\ket}[1]{|#1\!>}
\begin{document}
\vspace*{3cm}
\begin{center}
{\LARGE AXIOMATIC FOUNDATIONS OF \\[-.009cm]NONRELATIVISTIC QUANTUM
MECHANICS: \\[.1cm] A REALISTIC APPROACH $^{\dagger}$}\\[.9cm]
S. E. Perez  Bergliaffa$^{*}$ and H. Vucetich$^{*}$ \\
{\sl Departamento de F\'{\i}sica, UNLP, C.C. 67, La Plata, C.P. 1900,
Argentina}\\[.1cm] and \\[.1cm] G. E. Romero$^{*}$ \\ {\sl Instituto
Argentino de Radioastronom\'{\i}a, C.C. 5, Villa Elisa, C.P.
1894,\\ Argentina} \\[.2cm] December, 1993 \\[.4cm]
\begin{abstract}
A realistic axiomatic formulation of nonrelativistic quantum mechanics for
a single microsystem with spin is presented, from which the most important
theorems of the theory can be deduced. In comparison with previous
formulations, the formal aspect has been improved by the use of certain
mathematical theories, such as the theory of equipped spaces, and group
theory. The standard formalism is naturally obtained from the latter, starting
from a central primitive concept: the Galilei group.
\end{abstract}
\end{center}
\vspace*{\fill}
\hs $^{\dagger}$ Published in International Journal of Theoretical Physics,
{\bf 32}, 9, 1993.
\\ \indent $^{*}$ CONICET
\\ \indent {\sl E-mail: bergliaffa@fisilp.edu.ar}
\section*{INTRODUCTION}
\hs Every physical theory is a hypothetical-deductive system. This
system can be presented in several different ways. In our opinion, the
axiomatic approach has plenty
of advantages when compared with others. Firstly, in an axiomatization,
all the presuppositions
of the theory are explicited. This is very important to clarify the
foundations of the theory. Secondly, there is no place for doubts about the
arguments of the functions that appear in the statements. In this way, possible
erroneous identification of the physical referents can be avoided. Thirdly,
the meanings are assigned by semantical axioms, and not by context. This
excludes the frequent mistakes originating in an abuse of analogy. Finally,
the axiomatic formulation paves the way to the deduction of new theorems
and the elimination of pseudotheorems, because it clarifies the structure
of the theory.\\
\indent The proliferation of interpretations in the case of Quantum Mechanics
(QM) is partially due to semantical confusions arising from the
non-explicitation of the presuppositions. The standard axiomatization of QM
{\cite{von}} has
semantical contradictions, because it contains predicates that are not related
to the primitives that constitute the basis of the theory {\cite{bu1},
\cite{bu2}}. Bunge
has carried out a realistic axiomatization of QM, from which it is possible
to deduce the standard theorems of the nonrelativistic theory
{\cite{bu1},\cite{bu3}} . The present
work is a reformulation of that axiomatization which includes
several improvements. More specifically: \\
\begin{enumerate}
\item Our axiomatization has been formulated in an abstract way, in the sense
that it does not depend on any
particular representation, and presents the Schr\"{o}dinger equation,
the Heisenberg equation, and the hamiltonian of a free microsystem as
theorems.\\
\item The use of group theory enhaces the role played by symmetries in QM.\\
\item The mass and the charge have been eliminated from the generating basis
(which has in consequence been reduced). Both properties are introduced by
means of operators. \\
\item The spin is brought out directly from the rotational symmetry of the
system. \\
\item The theory of generalized functions developed by Gel'fand and Shilov
enables us to treat all the operators on an equal footing by the use of the
equipped Hilbert space.\\
\item Bargmann's superselection rules are presented as theorems.\\
\end{enumerate}
\hs In spite of the formal changes, the reference class of the theory
remains the same.
This implies that the interpretation of our formulation is realistic
 and literal.\\
\indent In the first section of this work, we describe some tools to
be used in the axiomatization. In the second section we give our axiomatic
formulation: background,
definitions, axioms and theorems. In the third section, we briefly discuss
some semantical aspects, and finally, in the last section, the
conclusions.
\section{TOOLS}
\hs We give next some mathematical and physical concepts that will be used
in the axiomatic core of the theory.
\subsection{THE GALILEI GROUP}
\hs The proper Galilei group {\cite{bar},\cite{lev}} contains the temporal
and spatial translations, the pure Galilei transformations, and the spatial
rotations. A general element of the group has the form	\\
\be
g=(\tau,\;\vec{a},\;\vec{v},\;R)
\ee
where $\tau$ is a real number, $\vec{a}$ and $\vec{v}$ are arbitrary vectors,
and $R$ is an orthogonal transformation. If $\vec{x}$ is a position vector
and $t$ is the time, a transformation belonging to the Galilei group is
\begin{eqnarray*}
\vec{x'} & = & R\vec{x}+\vec{v}t+\vec{a}\\
t' & = & t+\tau
\end{eqnarray*}
\indent The multiplication law is given by
\begin{eqnarray*}
g_{1}g_{2} & = & (\tau_{1},\vec{a_{1}},\vec{v_{1}},R_{1})(\tau_{2},
\vec{a_{2}},\vec{v_{2}},R_{2}) \\
 & = & (\tau_{1}+\tau_{2},\vec{a_{1}}+R_{1}\vec{a_{2}}
+\tau_{2}\vec{v_{1}},\vec{v_{1}}+R_{1}\vec{v_{2}},R_{1}R_{2})
\end{eqnarray*}
The unit element of the group is  \\
\be
e=(0,0,0,1)
\ee
and the inverse element of g is   \\
\be
g^{-1}=(-\tau,-R^{-1}(\vec{a}-\tau\vec{v}),-R^{-1}\vec{v},R^{-1})
\ee
\indent In\"{o}nu and Wigner {\cite{ino}} have shown that the basis
functions of the
representations of the Galilei group cannot be interpreted as wave functions of
physical microsystems, because it is impossible to construct well-localized
states or states with a definite velocity with them. Moreover, Hamermesh
{\cite{ham}} pointed out that the position operator can only be constructed
in the case of nontrivial ray representations.\\
\indent Bargmann {\cite{bar}} has shown that the physical representations of
the
Galilei group are obtained from the unitary ray representations of the
universal
covering group of the Galilei group. The exponents of these physical
representations have the form \\
\be
\xi(\overline{g_{1}},\overline{g_{2}})=\frac{1}{2}\{\vec{a_{1}}\cdot
R_{1}\vec{v_{2}}-\vec{v_{1}}\cdot R_{1}\vec{a_{2}}+\tau_{2}\vec{v_{1}}\cdot
R_{1}\vec{v_{2}}\}
\ee
where $\overline{g_{1}}=(\tau_{1},\vec{a_{1}},\vec{v_{1}},R_{1})$,
$\overline{g_{2}}=(\tau_{2},\vec{a_{2}},\vec{v_{2}},R_{2})$ are elements
of the universal covering group. To these elements correspond the unitary
operators $\hat{U}(\overline{g_{1}})$ and $\hat{U}(\overline{g_{2}})$ such that
\be
\hat{U}(\overline{g_{1}})\hat{U}(\overline{g_{2}})=e^{i\xi(
\overline{g_{1}},\overline{g_{2}})}\hat{U}(\overline{g_{1}},\overline{g_{2}})
\ee
\indent It is possible to construct a local group $\tilde{G}$ in the form
\be
\tilde{G}=(\theta,\overline{G})
\ee
where $\theta\in\cal R$ and $\overline{G}$ is the universal covering
group of the Galilei group $G$. We say that $\tilde{G}$ is a nontrivial
central
extension of the universal covering group $\overline{G}$ of the Galilei
group $G$ by a one-dimensional abelian group.\\
\indent The structure of $\overline{G}$ is locally determined by the structure
of its Lie algebra. The commutation relations among the elements of the
basis of the algebra can be calculated from the composition laws of $G$. For
the generators of spatial translations ($\hat{P}_{i}$) {\bf [1]}
and the
generators of pure Galilei transformations ($\hat{K}_{i}$), the
commutator is identically zero; if we compute this commutator
for the elements of the physical representation {\cite{lev}} we get
$[\hat{K}_{i},\hat{P}_{j}]=\hat{M}\delta_{ij}$. We carry out the central
extension of $\overline{G}$ imposing this last relation, in such
a way that $\hat{M}$ is the element of the Lie algebra of the one-parameter
subgroup used in the extension. This extension is central because $\hat{M}$
commutes with all the other elements of the algebra, and it is nontrivial
because $\hat{M}$ appears on the right side of some commutation relations.
The physical representations are then the representations of the algebra of the
central extension of $\ove{G}$.\\
\indent In our axiomatic formulation, the commutation relations
of the algebra of $\tilde{G}$ are explicitly postulated, and the generator
of the algebra of the one-parameter subgroup is identificated with the mass
operator $\hat{M}$.\\
\indent Let's turn now to the equipped Hilbert spaces and Gel'fand's
theorem.

\subsection{EQUIPPED HILBERT SPACES}
\hs As is well known, not all the physically important operators
appearing
in QM have eigenfunctions with finite norm. That is the case of the position
operator $\hat{X}$ and the linear momentum operator $\hat{P}$. In a
consistent axiomatic frame, all the eigenfunctions of operators
associated to physical properties should belong to a common space. The Hilbert
space $\cal H$ contains only normed vectors. It is then necessary to introduce
an extension: the equipped Hilbert space ${\cal H}_{e}$. This is not really
a space, but a 3-ple, given by
\be
{\cal H}_{e}=<\cal S,\;\cal H,\;\cal S'>
\ee
where $\cal S$ is a nuclear countable Hilbert space {\cite{gelu}} (i.e. a
space of well-behaved functions), $\cal H$ is the ordinary Hilbert space,
and $\cal S'$ is a space isomorphic to the dual of $\cal S$ (the distributions,
such as Dirac's delta, are in $\cal S'$). These three spaces satisfy
\be
{\cal S}\subset{\cal H}\subset{\cal S'}
\ee
\indent The following theorem, due to Gel'fand {\cite{gelu}}, that we
reproduce
without proof, states the necessary conditions to operate on $\cal S'$
in the usual way:\\
``Let ${\cal H}_{e}=<{\cal S},\;{\cal H},\;\cal S'>$ be an equipped
Hilbert space. If the symmetric and linear operator $\hat{A}$ acting on
the space $\cal S$ admits a self-adjoint extension $\bf\ove{A}$ on $\cal
H$, then $\bf\ove{A}$ admits a complete system of eigendistributions
$\{e_{r}\}$ in $\cal S'$ with real eigenvalues''.\\

We now define the action of the operators $\hat{X}$ and $\hat{P}$ on $\cal
S$ (in the corresponding representation) in the following way:
\be
\hat{X}\phi_{r}(x)=x\phi_{r}(x)
\ee
\be
\hat{P}\phi_{r}(p)=p\phi_{r}(p)
\ee
where $\{\phi_{r}\}$ is a complete set {\bf [2]}
The extension
of the operators (required by the theorem) can be achieved following Gel'fand
and Shilov {\cite{gelu}}.\\
\indent Gel'fand's theorem then legalizes the use of eigenfunctions of infinite
norm within the formal structure of the theory.

\section{AXIOMATICS}
\hs Every axiomatic formulation must explicit its background (i.e. the set
of all presuppositions of the theory), its
basis of primitive concepts (i.e. the set of non-definite concepts that
define the derived concepts according to the building rules explicited in
the background), its axioms, and its conventions. \\
\indent There are three kinds of axioms in a theory: formal axioms,
physical axioms, and semantical axioms.
The formal axioms are of a purely mathematical type and they refer only
to conceptual objects. The physical axioms represent objective physical
laws. The semantical axioms establish the relations among signs, physical
objects and properties of physical objects; in this way they
characterize
the meaning of the primitives and they set the reference class of the theory.\\
\indent We next give the background of our formulation of nonrelativistic
quantum mechanics for one microsystem (T$_{QM}$).

\subsection{FORMAL BACKGROUND}
\lista{Pr}{P}
\item Two-valued ordinary logic
\item Formal semantics {\cite{bu4}\cite{bu5}}
\item Mathematical analysis with its presuppositions and generalized functions
theory {\cite{gel}\\ \cite{jon}}
\item Probability theory
\item Group theory
\end{list}
\subsection{MATERIAL BACKGROUND}
\lista{Prd}{P}
\setcounter{Prd}{5}
\item Chronology
\item Euclidean physical geometry {\cite{bu1}}
\item Physical theory of probability {\cite{pop}}
\item Dimensional analysis
\item Systems theory
\item Classical electrodynamics
\end{list}
\subsection{REMARKS}
\hs By chronology we understand the set of theories of time. We adopt
here a relational theory for the local time, in which a function is defined
such that it maps pairs of events related to a given reference system into
a segment of the real line {\cite{bu1}}. \\
\indent The theory of systems deals with physical systems and the
relations among them (a  physical system is ``\ldots anything existing in
space-time and such that it either behaves or is handled as a whole in at least
one respect'' {\cite{bu1}}). This theory has been axiomatized by Bunge
{\cite{bu1}},
and its basis of primitive concepts includes the physical sum or juxtaposition
( $\dot{+}$ ), and the physical product or superposition ( $\dot{\times}$ ).\\
\indent Finally, the inclusion of classical electrodynamics will allow , by
means of the axiom $\bf A_{42}$, the study a microsystem under the
influence
of an external classical field. The removal of $\bf P_{11}$ causes the axioms
$\bf A_{37},\;\bf A_{38},\;$ and $\bf A_{42}$ to be meaningless. \\
\indent Let's turn now to the generating basis.

\subsection{GENERATING BASIS}
\hs The conceptual space of the theory is generated by the basis B of
primitive concepts, where \\
\begin{center}
B=$\{\Si,\:\Sir,\:$E$_{3},\:$T$,\:{\cal H}_{e},\:{\cal P},\:$A, G, $\hbar\}$
\end{center}
\indent The elements of the basis will be semantically interpreted by means of
the axiomatic basis of the theory, with the help of some conventions
{\bf [3]}.
\subsection{DEFINITIONS}
\lista{De}{D}
\item $eiv\:\hat{A}=_{Df}$ eigenvalue of  $\hat{A}$
\item $[\hat{A},\hat{B}]=_{Df}\;\hat{A}\hat{B}-\hat{B}\hat{A}$
\item $\Psi=_{Df}\:\{\alpha |\psi_{0}\!>\::(\alpha\:\in\cal C,$ with $|\alpha|
=1)\:\wedge(|\psi_{0}\!>\;\in\cal H$ is a fixed vector)\} is a ray in $\cal H$
\item If $|\psi\!>\;\in\Psi\:\subset\:{\cal H}\:\Rightarrow\:|\psi\!>$ is
a representative of $\Psi$
\item If the spectrum of $\hat{A}$ is continuous $\Rightarrow \\
<\!\psi|\hat{A}|\phi\!>=_{Df}\:\int da\:db\:<\!\psi|a\!><\!a|\hat{A}|b\!>
<\!b|\phi\!>=\int da\:db\;\psi^{*}(a)\:A_{ab}\:\phi(b)$
\item If the spectrum of $\hat{A}$ is discrete $\Rightarrow\\
<\!\psi|\hat{A}|\phi\!>=_{Df}\:\sum_{i,j}\:<\!\psi\ket{a_{i}}<\!a_{i}|\hat{A}\ket{b_{j}}<\!b_{j}\ket{\phi}=
\;\sum_{i,j}\:\psi_{i}\:A_{ij}\:\phi_{j} $
\item $\Psi.\Phi=_{Df}|<\!\psi|\phi\!>|$
\item ${\cal U}=_{Df}\{\alpha\hat{U}_{0}\::(\alpha\;\in \cal C$, with
$|\alpha|=1)\:\wedge\:(\hat{U}_{0}$ is a fixed unitary operator on $\cal H)\}$
is a ray operator on $\cal H$
\item If $\hat{U}\:\in{\cal U}\:\Rightarrow\hat{U}$ is a representative
of  $\:\cal U$
\item If ($\ket{\psi}\;\in\;\Psi)\;\wedge(\ket{\psi '}\;\in\;\Psi)\;\wedge\;
(\ket{\psi '}=e^{i\theta}\:\ket{\psi})\Rightarrow
\:\ket{\psi '}=_{Df}$ gauge transformed by a gauge transformation of the first
kind of $\ket{\psi}$.
\end{list}
\subsection{AXIOMATIC BASIS}
\hs T$_{QM}$ is a finite-axiomatizable theory, whose axiomatic basis is
{\cite{bu4}}
\begin{center}
${\cal B}_{A}(T_{QM})=\bigwedge^{42}_{i=1}$ {\bf A$_{i}$}
\end{center}
where the index $i$ runs over the axioms.
\subsection{AXIOMS}
\subsubsection*{GROUP I: SPACE AND TIME}
\lista{Ax}{A}
\item E$_{3}\equiv$ tridimensional euclidean space.
\item E$_{3}\:\hateq$ physical space.	\sa
\item T$\:\equiv$ interval of the real line R.
\item T$\:\hateq$ time interval.   \sa
\item The relation $\leq$ that orders T means ``before to'' $\vee$
``simultaneous with''.   \sa
\subsubsection*{GROUP II: MICROSYSTEMS AND STATES}
\setcounter{Ax}{5}
\item $\Si ,\ove{\Si}$: non-empty, denumerable sets.
\item $\forall \si \in \Si, \si$ denotes a microsystem. In particular,
$\si _{0}$ denotes absence of microsystem.   \sa
\item $\forall\:\sir \in \Sir,\:\sir$ denotes environment of some system. In
particular, $\sirce$ denotes the empty environment, $\cero$ denotes a free
microsystem, and $<\!\si _{0},\ove{\si} _{0}\!>$ denotes the vacuum.   \sa
\item $\forall \sig\;\in \Sig ,\:\exists \:{\cal H}_{e} \ni {\cal H}_{e}
=<{\cal S},\;{\cal H},\;{\cal S'}>\:\equiv$ equipped Hilbert space.
\item There exists a one-to-one correspondence between physical states of
$\si\in
\Si$  and rays $\Psi\subset\:\cal H$.	\sa
\subsubsection*{GROUP III: OPERATORS AND PHYSICAL QUANTITIES}
\setcounter{Ax}{10}
\item $\cal P\equiv$ non-empty family of functions on $\Si$.
\item A $\equiv$ ring of operators on ${\cal H}_{e}$.
\item $\forall\:\cal A\:\in\cal P,\cal A$ designates a property of
$\si\:\in\Si$
.  \sa
\item ($\forall\:{\cal A}\in\:{\cal P})\:\exists\:
\hat{A}\:\in\:$A$\:\ni\hat{A}\:\hateq{\cal A}$.   \sa
\item (Hermiticity and
linearity)\\($\forall\:\si\in\Si)\wedge
\:\ni t_{0}$ is fixed) $\wedge\;(\forall\hat{A}\:\in\:$A$\:\ni\hat{A}
\hateq{\cal A},\;{\cal A}\;\in{\cal P}$) if $|\psi_{1}\!>,\:|\psi_{2}\!>\:\in
{\cal H}_{e}\Rightarrow$
\begin{enumerate}
\item $\hat{A}:{\cal H}_{e}\rightarrow{\cal H}_{e}\:\ni\hat{A}[\lambda_{1}
|\psi_{1}\!>+\lambda_{2}|\psi_{2}\!>]=\lambda_{1}\hat{A}|\psi_{1}\!>\!
+\lambda_{2}\hat{A}|\psi_{2}\!>$ with $\:\lambda_{1},\lambda_{2}\in\:\cal C$
\item $\hat{A}^{\dagger}=\hat{A}$ on $\cal H$.
\end{enumerate}
\item
(Probability densities)\\
($\forall\sig\:\in\Sig)\wedge\;(\forall\hat{A}\;\in\:$A$\:\ni\hat{A}
\hateq{\cal A},\:{\cal A}\:\in{\cal P})\wedge\:(\forall\:|a\!>\:\in{\cal H}\:
\ni\hat{A}|a\!>=a|a\!>)\wedge\:
(\forall\:|\psi\!>\:\in\Psi\:\subset\cal H$ that corresponds to the
state of $\si$ when it is influenced by $\sir$):

$<\!\psi|a\!><a|\psi\!>\equiv$ probability density for the property $\cal
A$ when $\si$ is associated to $\sir$\\ (i.e. $\int^{a_{2}}_{a_{1}}
<\!\psi|a\!><\!a|\psi\!>\;da$ is the probability for $\si$ to have an
$\cal A$-value in $[a_{1},a_{2}]$).   \sa
\item ($\forall\:\si\:\in\Si)\:\wedge(\forall\:\sir\:\in\Sir$) the ray $\Psi$
corresponding to a state of $\si$ is the null ray on the border of the
accesible region for the system $\si\dot{+}\sir$.
\item ($\forall\:\si\in\Si)\:\wedge(\forall\hat{A}\:\in$ A
$)\wedge(\fa a\:\ni eiv\:\hat{A}=a) \:a$
is the sole value that $\cal A$ takes on $\si$, given that $\hat{A}
\hateq\cal A$.	 \sa
\item $\hbar\:\in\cal R^{+}$.
\item $[\hbar]=LMT^{-1}$.
\subsubsection*{GROUP IV: SYMMETRIES AND GROUP STRUCTURE}
\setcounter{Ax}{20}
\item (Unitary
operators)\\($\forall\sig\:\in\Sig)\:\wedge(\forall\hat{A}\:\in\:
$A$\:\ni\hat{A}\:\hateq\cal A,\:\cal A\:\in\cal P$) if $\exists\:\hat{U}
\ni\hat{U}^{\dagger}=\hat{U}^{-1}\:\Rightarrow\hat{A}'=\hat{U}^{\dagger}\hat{A}
\hat{U}\:\hateq\cal A$.   \sa
\item $\forall\:\cero\:\in\Sig \;\exists\;\hat{D}(\tilde{G})$, unitary
ray
representation of some central non-trivial extension of the universal
covering group $\bar{G}$ of a Lie group $G$ by a one-dimensional abelian
group on $\cal H$.
\item The Lie algebra $\cal G$ of the group $G$ is generated by
\{$\hat{H},\;\hat{P_{i}},\;\hat{K_{i}},\;\hat{J_{i}}$\}$\:\subset A$.
\item (Algebra structure)\\The structure of $\tilde{\cal G}$, Lie
algebra of $\tilde{G}$ is: \\ [.5cm]
$\conu{J}{i}{J}{j}{k}{J}\hfill\conu{J}{i}{K}{j}{k}{K}\hfill\conu{J}{i}{P}{j}{k}
{P}$ \\[.5cm]
\hspace*{\fill} $[\hat{K}_{i},\hat{H}]=i\hbar\hat{P}_{i} $\hfill
$[\hat{K}_{i},\hat{P}_{j}]=i\hbar\delta_{ij}\hat{M}\hfill $
\begin{center}
$\begin{array}{cccc}
\cont{J}{i}{H}	 &\cond{K}{i}{K}{j}  &\cond{P}{i}{P}{j} &\cont{P}{j}{H} \\[.3cm]
\cont{J}{i}{M}	 &\cont{K}{i}{M}     &\cont{P}{i}{M}	&[\hat{H},\hat{M}]=0
\end{array}$
\end{center}
where $\hat{M}$ is an element of the Lie algebra of a one-parameter subgroup
(which is used to extend $\ove{G}$).
\item $G$ is the Galilei group.
\item $\hat{H}$ is the time-translations generator.
\item $\forall\sig\:\in\Sig,\:eiv\:\hat{H}=E$ represents the energy value
of $\si$ when it is influenced by $\sir$.   \sa
\item $\op{P}{i}$ is the generator of spatial translations on the cartesian
coordinate axis $X_{i}$.
\item $\forall\sig\:\in\Sig,\:eiv\:\op{P}{i}=p_{i}$ represents the
$i$-component of the linear momentum of $\si$.	 \sa
\item $\op{J}{i}$ is the generator of spatial rotations around the
cartesian coordinate axis $X_{i}$.
\item $\forall\:\sig\in\:\Sig,\:eiv\:\op{J}{i}=j_{i}$ represents the
$i$-component of the angular momentum of $\si$.   \sa
\item $\op{K}{i}$ is the generator of pure transformations of Galilei
on the axis $X_{i}$.
\item $\hat{M}$ has a discrete spectrum of real and positive eigenvalues.
\item $\forall\sig\:\in\Sig,\:eiv\:\hat{M}=\mu$ represents the mass of $\si$.
\sa
\item $\forall\sig\:\in\Sig$, if $\op{X}{i}=_{Df}\frac{1}{\mu}\op{K}{i}$, then
$eiv\:\op{X}{i}=x_{i}$ represents the $i$-component of the position of
$\si$.
  \sa
\subsubsection*{GROUP V: GAUGE TRANSFORMATIONS AND ELECTRIC CHARGE}
\setcounter{Ax}{35}
\item$(\forall\sig\:\in\Sig)\:\exists\:\hat{Q}\in\:$A$\:\ni(\hat{Q}\neq
\hat{I})\wedge\:([\hat{Q},\hat{A}]=0\;\forall\hat{A}\in$ A).
\item $\hat{Q}$ has a discret spectrum of real eigenvalues.
\item $\hat{Q}$ is the generator of gauge transformations of the first kind.
\item $\forall\sig\:\in\Sig,\:eiv\:\hat{Q}=q$ represents the charge of $\si$.
  \sa
\item There exists one and only one normalized state with $eiv\:\hat{Q}=0$,
named the neutral state.
\item There exists one and only one normalizable state, named vacuum, that
is invariant under $\hat{D}(\tilde{G})$ and under gauge transformations of the
first kind.
\item If $\si\:\in\Si,\:eiv\:\hat{M}=\mu\neq 0,\:eiv\:\hat{Q}=e$ and
$<\!A_{0},\vec{A}\!>$
are the components of an electromagnetic quadripotential that represents
the action of $\sir\;\neq\sirce$ on $\si\;\Rightarrow$
\begin{center}
$\hat{H}=\frac{1}{2\mu}(\hat{\vec{P}}-\frac{e}{c}\vec{A})^{2}+
\frac{e}{c}\:A_{0}-g_{l}\frac{\hbar e}{mc}\vec{B}.\hat{\vec{\si}}$
\end{center}
where  $\vec{B}$ has the usual meaning that follows from $\bf P_{10}$,
$\hat{\vec{\si}}$ is specified in $\bf T_{13}$ and $g_{l}$ is the gyromagnetic
factor of the microsystem.
\end{list}
\subsection{REMARKS}
\hs From the axioms, it can be seen that the algebra $\cal S$ of the
symmetry group
S of T$_{QM}$ for $\cero\in\Sig$ consists of two ideals: an 11-dimensional
ideal corresponding to the central extension of the algebra of the universal
covering group of the Galilei group by a one-dimensional Lie algebra, and
a one-dimensional abelian ideal corresponding to the U(1) algebra, whose
generator is $\hat{Q}$. Stated matematically, S=$\tilde{G}\otimes$U(1).\\
\indent In the case of $\si\neq\sirce$, the group of symmetries will depend on
the explicit form of $\hat{H}$, and its algebra will be some subalgebra of
$\cal S$. \\
\indent The theorems will show that the physics is mainly contained in the
commutation relations given in $\bf A_{24}.$
\subsection{DEFINITIONS}
\lista{Df}{D}
\setcounter{Df}{10}
\item Non-degenerated spectrum of an operator $\hat{A}\ni\hat{A}\ket{\phi}=a
\ket{\phi}$ (with given boundary conditions) where  $\hat{A}\in$ A and
$\ket{\phi}\:\in\Phi\subset {\cal H}=_{Df}\{a\}\ni(\fa a\in\{a\}\:\exists\:
\ket{\phi}\in\{\ket{\phi}:\hat{A}\ket{\phi}=a\ket{\phi}\})\wedge(\{a\}\cong
\{\ket{\phi}\})$
\item Component of $\ket{\psi}$ along $\ket{\phi_{k}}=_{Df}<\!\phi_{k}|\psi\!>
=c_{k}$
\item $<\!\hat{A}\!>=_{Df}<\!\psi|\hat{A}\ket{\psi}$
\item $\Delta\hat{A}=_{Df}\hat{A}-<\!\hat{A}\!>$
\item $(\Delta\hat{A})^{2}=_{Df}<\!(\hat{A}-<\!\hat{A}>)^{2}\!>=<\!\hat{A}^{2}-
<\!\hat{A}\!>^{2}>$
\item $|\!|\psi|\!|^{2}=_{Df}<\!\psi|\:\psi>$
\item $\hat{S}_{i}=_{Df}\frac{\hbar}{2}\hat{\sigma}_{i}$
\item $\hat{L}_{i}=_{Df}\epsilon_{ijk}\hat{X}_{j}\hat{P}_{k}$
\item Time evolution operator$=_{Df}\teo{t}{t_{0}}\ni(\teo{t}{t_{0}}
\teo{t}{t_{0}}^{\dag}=\hat{I})\;\wedge\;(\teo{t}{t'}\teo{t'}{t_{0}}=
\teo{t}{t_{0}})\;\wedge\;(\teo{t_{0}}{t_{0}}=\hat{I})\;\wedge\;(\hat{A}(t)=
\teo{t}{t_{0}}^{\dag}\hat{A}(t_{0})\teo{t}{t_{0}})$
\end{list}
\subsection{THEOREMS}
\hs In this section we give some illustrative theorems that can be
deduced from the axioms.
\lista{Th}{T}
\item (Probability amplitudes)\\
The probability that the property ${\cal A}$ represented by a non-degenerate
operator $\hat{A}$ of the composed system $\si\dot{+}\sir$ in the state $\Psi$
takes a value $a_{k}\in\{a_{k_{1}},a_{k_{2}}\}$ is given by
\bc
$P(a_{k})=\sum_{k\in\Delta k}|c_{k}|^{2}$
\ec
\hfill$\Delta k=\{k_{1},k_{2}\}$
where $c_{k}=<\!\phi_{k}|\psi>$ and $\ket{\phi_{k}}$ is an eigenvector of
$\hat{A}$\\
Proof: see {\cite{bu1}}, p. 252.
\item Under the same conditions of {$\bf T_{1}$}, the average of
$\hat{A}$ is:
\bc
$<\!\hat{A}\!>=\sum_{k}|c_{k}|^{2}a_{k}$
\ec
Proof: from {$\bf P_{4}$} and {$\bf T_{1}$}.
\item ($\fa\sig\in\Sig)\wedge(\fa\hat{H}\neq\hat{H}(t)\ni\hat{H}$ is the
generator of temporal translations) the time evolution operator is:
\bc
$\teo{t}{t_{0}}=exp\{-\frac{i}{\hbar}\hat{H}(t-t_{0})\}$
\ec
Proof: using {$\bf A_{24}$ and $\bf A_{26}$}.
\item (Schr\"odinger equation)\\
If $\ket{\psi}_{t}=\teo{t}{t_{0}}\ket{\psi}_{t_{0}}\in\Psi$ is a representative
of the state of $\si\in\Si$ when $\si$ is influenced by $\sir\in\Sir$ then
$\ket{\psi}_{t}$ satisfies:
\bc
$\hat{H}\ket{\psi}_{t}=i\hbar\frac{\partial\ket{\psi}_{t}}{\partial t}$
\ec
Proof: from {$\bf T_{3}$}.
\item $\fa\cero\in\Sig,\;\hat{H}=\frac{\hat{P}^{2}}{2\mu}$  \\
Proof: from {$\bf A_{24}$} (see {\cite{ham}}).
\item $\fa\sig\in\Sig$ the properties ${\cal A},{\cal B}\in{\cal P}$ take
definite values at the same time if and only if the associated operators
$\hat{A}$ and $\hat{B}$ have the same eigenvectors. \\
Proof: using {$\bf D_{14}$}.
\item The operators $\hat{A}$ and $\hat{B}$ of $\bf T_{6}$ have a common
basis of eigenvectors if and only if they commute.   \\
Proof: using {$\bf D_{5}$}.
\item (Heisenberg's inequalities)\\
$(\fa\sig\:\in\Sig)\:\wedge\:(\:\fa\;\ket{\psi}\in{\cal H})\:\wedge\:
(\:\fa\;\{\hat{A},
\hat{B},\hat{C}\}
\subset$ A $\ni\hat{A}\hat{=}{\cal A},\;\hat{B}\hat{=}{\cal B},\; \hat{C}
\hat{=}{\cal C}$ with $\{{\cal A,\;\cal B,\;\cal C}\}\subset{\cal P}$) if
$[\hat{A},\;\hat{B}]=i\hat{C}\Rightarrow$
\bc
$(\Delta\hat{A})^{2}(\Delta\hat{B})^{2}\geq |\hat{C}|^{2}/4$
\ec
Proof: using {$\bf D_{12}$}, {$\bf D_{14}$}, Schwartz's inequality, and
the definition $\hat{F}=\hat{A}\hat{B}+\hat{B}\hat{A}$.
\vspace*{.5cm} \\
{\bf COROLLARY:} If [$\hat{X}_{i}, \hat{P}_{j}]=\:\hbar\:\delta_{ij}\:\hat{I}$
then \\
\bc
$\Delta\hat{X_{i}}\:\Delta\hat{P_{j}}\geq\hbar/2$
\ec
\item (Heisenberg's equation)\\
$(\fa\sig\in\Sig)\wedge(\fa\hat{A}\in$ A  $\ni\hat{A}\hat{=}{\cal A}$, $\cal A
\in\cal P$):
\bc
$\frac{d\hat{A}}{dt}=\frac{i}{\hbar}[\hat{H},\;\hat{A}]$
\ec
Proof: from {$\bf D_{18}$} and {$\bf T_{3}$}.\vspace*{.5cm} \\
{\bf COROLLARY}: if $[\hat{H},\:\hat{A}]=0\Rightarrow\hat{A}$ represents
a constant of motion.
\item ($\fa\sig\in\Sig)\wedge(\fa\:\ket{\psi}\in{\cal H})\wedge
(\fa\hat{A} \in$ A $\ni\hat{A}
\hat{=}{\cal A}$ with ${\cal A}\in{\cal P})\wedge(\fa\hat{H}\ni[\hat{H},\;
\hat{A}]=i\hat{C}$):
\bc
$\Delta\hat{H}\:\tau_{A}\geq\frac{\hbar}{2}$
\ec
with $\tau_{A}=\Delta\hat{A}/|d<\!\hat{A}\!>/dt|$.    \\
Proof: from {$\bf D_{12}$}, {$\bf T_{8}$} and {$\bf T_{9}$}.
\item If $\op{J}{i}$ is the spatial rotations generator around the axis
$x_{i}$ $\Rightarrow$ \\ \hfill $[\hat{J}^{2},\;\op{J}{i}]=0$. \hfill
\\ Proof: using {$\bf A_{24}$}.
\item If $\ket{j,\;m}$ is an eigenstate of $\hat{J}^{2}$ and $\op{J}{3}$
then
\bc
$\hat{J}^{2}\ket{j,\;m}=\hbar^{2}j(j+1)\ket{j,\;m}$\\
$\op{J}{3}\ket{j,\;m}=\hbar m\ket{j,\;m}$
\ec
with $-j\leq m\leq -m,\;j$ half-integer.\\
Proof: from {$\bf A_{24}$} and {$\bf T_{11}$}, using
$\hat{J}_{\pm}=\hat{J}_{1} \pm i\hat{J}_{2}$.

\item (Spin) \\
If $j=1/2\Rightarrow\hat{\vec{J}}=(\op{J}{1},\;\op{J}{2},\;\op{J}{3})=
\frac{\hbar}{2}\hat{\vec{\si}}$, with
$\hat{\vec{\si}}=(\op{\si}{1},\;\op{\si}{2},\;\op{\si}{3})$, and
\bc
$\si_{1}=\left( \begin{array}{cc}
0 & 1\\
1 & 0
\end{array}\right)\;\;
\si_{2}=\left( \begin{array}{cc}
0 & -i \\
i & 0
\end{array}\right)\;\;
\si_{3}=\left( \begin{array}{cc}
1 & 0 \\
0 & -1
\end{array}\right)$
\ec
Proof: from {$\bf D_{5}$} and {$\bf T_{12}$}, using $\hat{J}_{\pm}$.
\item $\fa\sig\in\Sig,\;\hat{\vec{J}}=\hat{\vec{L}}+\hat{\vec{S}}$  \\
Proof: from {$\bf D_{17}$}, {$\bf D_{18}$} and {$\bf A_{24}$}.
\item (Superselection rules)\\
$\fa\sig\in\Sig,\;{\cal H}$ decomposes in mutually orthogonal subspaces
whose vectors are eigenvectors of $\hat{M}$. The same is valid for the charge
operator $\hat{Q}$.    \\
Proof: from {$\bf A_{24}$} and {$\bf A_{36}$}.
\end{list}
\subsection{REMARKS}
The theorem {$\bf T_{5}$} gives the form of $\hat{H}$ for a free
microsystem;
its expression is deduced from the symmetry group (i.e. the Galilei group). The
time-translations generator $\hat{H}$ characterizes the Schr\"odinger's
equation
({$\bf T_{4}$}), which in turn enables us to calculate the vectors
corresponding to the physical states of the system. It is clear then that
the fundamental physical features of the theory are contained in {$\bf
A_{24}$}.
\\ \indent The theorem {$\bf T_{10}$} should not be taken as the
so-called
fourth Heisenberg's inequality: $\Delta E \Delta t\geq\hbar/2$, which is
meaningless in our formulation. In fact, being $t$ a parameter and not an
operator, this latter relation is not a logical consequence of {$\bf
T_{8}$}.
In the expression given in {$\bf T_{10}$} only the characteristic
time of the statistical evolution of the operator $\hat{A}$ (i.e. $\tau_{A}$)
appears.\\
\indent The superselection rule ({$\bf T_{15}$}) for the mass operator
$\hat{M}$
implies the conservation of the microsystem's mass in the processes that
can be described within our axiomatic frame (i.e those non-relativistic
processes that reduce to a problem involving a microsystem and its environment)
. This restriction also holds in Galilean Quantum Field Theories: it forbides
certain reactions in which annihilation and creation of particles occur
{\cite{lev2}}. Note that the superselection rule for the mass is a direct
consequence of the imposition of physical representations to the Galilei
group. In contrast, the corresponding rule for the charge must be presented
in a separated axiom.
\section{SEMANTICAL ASPECTS}
\hs The semantical structure of the theory is determined by the semantical
rules
expressed in the axioms (SA). This set of axioms fixes the factual
interpretation of the mathematical formalism, giving the theory a physical
status. \\
\indent The semantical axioms are of two kinds: denotation rules (like
$\bf A_{7}$
or $\bf A_{8}$) that relate symbols and referents in a conventional way,
and representation rules (like $\bf A_{14}$ or $\bf A_{21}$) that set
correspondences between functions (or other conceptual objects) and properties
of referents. These last rules are not conventional. Moreover, they are
hypothesis that can be empirically and theoretically contrasted {\cite{bu4}}.
This fact permits the discussion of the foundations of the theory, giving
to the variety of presented hypothesis, a variety of rival interpretations.
However, in most of the interpretations the semantical axioms are not clearly
identified from the rest of the axioms. This facilitates the propagation
of interpretation mistakes.\\
\indent A semantical axiom that usually appears in the standard formulation
of the theory is the so-called von Neumman's projection postulate: \\
``If the measurement of a physical observable $\cal A$ (with associated
operator
$\hat{A}$) on a quantum system in the state $\ket{\psi}$ gives a real value
$a_{n}$, then, inmediately after the measure, the system evolves from the state
$\ket{n}$, where $\hat{A}\ket{n}=a_{n}\ket{n}$''.\\
\indent This postulate interprets the collapse of the wave function as a
consequence
of the act of measuring the property $\cal A$. In our formulation of T$_{QM}$
this postulate plays no role. Morover, it is in contradiction with the rest
of the axiomatic core: neither the observer nor the measuring apparatus
are present in the background or the generating basis. As a consequence,
none of the legitimate statements in the theory can refer to them. Our
formulation is objective, realistic and literal. The microsystem-apparatus
interaction must be studied by the quantum theory of measurement, and there
are reasons {\cite{zin}} to think that also in this theory
the postulate in question can be eliminated.\\
\indent If T$_{QM}$ does not say anything about observers and measurements,
which is the kind of entities whose existence is presupposed by it?. To
ask this is to ask for the ontology of the theory. Vaguely, the ontology
is the answer, given by a theory,  to the question ``what is
there?''. More
precisely, we understand  the ontology in the following restricted sense:
the ontology of the theory is the factual restriction of the set formed
by the union of the domains of all the variables related to logical
quantifiers
that appear in the axiomatic basis of the theory (by factual restriction we
understand a restriction of the domain to the subset formed by all the non-
conceptual elements) {\bf [4]}
. In the axioms, we
quantify on
the elements of the generatig basis or on conceptual objects generated
by it. All the non-conceptual objects of ${\cal B}_{A}$ belong to $\Si\bigcup
\Sir$. That is why we identify this set with the ontology of T$_{QM}$. In
our restricted sense, the ontology coincides with the reference class of
the theory {\cite{bu4}}:
\be
R_{F}(T_{QM})=\bigcup^{42}_{i=1}R_{F}({\bf A_{i}})= \Si\cup\Sir
\ee
T$_{QM}$ refers then only to microsystems and its physical environments.
\section{CONCLUSIONS}
\hs We have presented in this work an axiomatization of nonrelativistic
quantum mechanics which displays without ambiguity the logical structure of
the theory, in such a way that any proposition is either a postulate or a
logical consequence of the postulates. In this form, there is no place for
statements unrelated to the primitive concepts of the generating basis.
Besides, the semantical structure of the theory has been totally explicited.
This avoids possible mistakes in the assignation of meaning to the different
symbols. These aspects, joined with the formal advantages mentioned in the
introduction, enable us to build an axiomatic picture of QM with a realistic
and objective interpretation.\\
\indent It is widely known that, in the subjective interpretations of QM, the
state
vector gives a complete characterization of only one microsystem. On the other
hand, in the realistic statistical interpretations, the state vector describes
an ensemble of microsystems. In the interpretation of the axiomatic basis here
presented, a ray in a Hilbert space characterizes a single microsystem in a
realistic (nonsubjectivistic) way.\\
\indent The formal structure developed in the present article is apt to
study a single microsystem with spin, with or without an external
electromagnetic field, and every problem that can be reduced to a one body
problem (e.g. hydrogen-like atoms). This limitation will be removed in a
future paper, which will generalize the axiomatic frame in such a way
that it will encompass the case of a microsystem with an arbitrary
number of components.
Particular attention will be payed to the symmetrization postulate and the
EPR paradox.
\subsection*{ACKNOWLEDGMENTS}
\hs We are grateful to M. Rocca and P. Sisterna for some helpful
discussions, and
specially to M. Bunge for valuable comments on the preparation of the
manuscript. 
\subsection*{Notes}
{\bf [1]} Latin indices can take the values 1, 2, 3.\\
{\bf [2]} It may seem that these definitions are somehow restrictive
because they depend on the corresponding representations. However, it is
shown in \cite{jon} that there exists an isomorphism between the $x$ and $p$
representations of $\cal S$ and $\cal S '$, whereas the isomorphism between
the representations of $\cal H$ can be deduced from Parseval's theorem. The
existence of these isomorphisms guarantees independence from any particular
representation. \\
{\bf [3]} We use an informal notation (with the risk of commiting
language abuses) instead of exact logical notation that would obscure
the physics of the problem.\\Some unusual symbols and their meaning:
$\hateq$ (``\ldots represents\ldots ''), $\ni$ (``\ldots such that\ldots ''),
$\tilde{=}$ (``\ldots isomorphic to\ldots '').  \\
{\bf [4]} In a strict sense, it should be understood that a X-logy is
a theory of X, for all X, and not just a set. However, Quine and others use
this
word in a different sense, related to the set of entities accepted by the
theory.
A more precise definition of this last acception is given here.
\end{document}